\documentclass[12pt]{article}
\usepackage{epsfig}
\begin{document}
\begin{center}
{\Large {\bf Measure in the 2D Regge quantum gravity. \\

}} \vskip-40mm \rightline{\small ITEP-LAT/2005-03} \vskip 30mm

{%\baselineskip=16pt
\vspace{1cm}
{ M.A.~Zubkov$^a$  }\\

 \vspace{.5cm} { \it
$^a$ ITEP, B.Cheremushkinskaya 25, Moscow, 117259, Russia }}
\end{center}
\begin{abstract}
We propose a version of the $2D$ Regge calculus, in which the areas of all
triangles are equal to each other. In this discretization Lund - Regge measure
over link lengths is simplified considerably. Contrary to the usual Regge
models with Lund - Regge measure, where this measure is nonlocal and rather
complicated, the models based on our approach can be investigated using the
numerical simulations in a rather simple way.
\end{abstract}

%\vfill
\today \hspace{1ex}
%\eject

\newpage
One of the main difficulties in quantization of gravity is the wideness of its
gauge group, that is the group of general coordinate transformations. Due to
this wideness it is not possible to put the theory into a regular lattice and
express it in terms of metrics without loss of gauge invariance. At the present
moment the only gauge invariant discretization of quantum gravity is Regge
calculus. However, the price for the explicit gauge invariance is the problems
with the definition of measure. Here instead of metrics, which was local field
in continuum, the lengths of the links are fundamental gauge invariant
dynamical variables. Metrics is composed of these variables in a nonlocal way.
Therefore, measure is also expected to be nonlocal. Although in \cite{W} and
related publications there were argued that certain choices of local measures
over link lengths can survive as appropriate, the point of view of the author
of the present Letter is different. Namely, we accept that the correct measure
over link lengths must be constructed in the spirit of construction of the so -
called Lund - Regge measure (see, for example, \cite{A1}, where it is also
called De - Witt like measure). As was mentioned above, this measure in general
case is nonlocal. Actually, it's form is so complicated, that it seems not
possible to use it in real numerical simulations.

However, in the present Letter we suggest the way to overcome this difficulty,
at least, in two dimensions. In order to do this we accept the compromise
decision. Namely, we start discretization of gravity from the partially gauge
fixed version of the theory, where the conformal mode is fixed via using
diffeomorphism invariance. As a result, we propose a version of the Regge
calculus, in which areas of all triangles are kept equal to each other. Within
this discretization of gravity Lund - Regge measure is simplified considerably,
which allows to perform real numerical investigation of the correspondent
models using Monte - Carlo methods.

To begin the description of our approach let us remind main facts about the
measure in quantum gravity theories.  First of all, what is the original space
of continuous geometries? We suppose here that our basic space to be disretized
is the set of all compact $D$ - dimensional Riemannian manyfolds of certain
implied smoothness properties. There is the following metrics on this set.
\begin{equation}
\|\delta g\|^2 = \frac{1}{2}\int d^D x \sqrt{|g|} (g^{\mu \nu}g^{\rho \eta} +
g^{\rho \nu}g^{\mu \eta} + C g^{\mu \rho}g^{\nu \eta})\delta g_{\mu \rho}
\delta g_{\nu \eta}\label{NORM}
\end{equation}

It is well known that in finite dimensional case any metrics (on the compact
space) produces the only measure $\mu$. In a few words the procedure of its
construction works as follows. First we set a certain $\epsilon$. Next, we put
balls of radius $\epsilon$ into the space. These balls should be putted in such
a way that they intersect each other only by their boundaries. They should be
arranged is such a way that their total number is maximal. After that for any
given set $\Omega \subset \Lambda$ (we denote the entire space by $\Lambda$) we
count the number of balls situated inside $\Omega$. Let us denote it as
$n_{\epsilon}(\Omega)$. Finally
\begin{equation}
\mu(\Omega) = {\rm lim}_{\epsilon \rightarrow 0}
\frac{n_{\epsilon}(\Omega)}{n_{\epsilon}(\Lambda)}
\end{equation}

In principle this procedure might be generalized to infinite dimensional case.
The spaces could be represented as a limit of finite dimensional
ones\footnote{This construction is, however, much more complicated and is not
yet elaborated in sufficient details for most cases of interest (at least the
author of the present Letter is not aware of the correspondent research).}.
Namely, each Riemannian manyfold could be thought of as a limit of a sequence
of piecewise linear manyfolds constructed of larger and larger number of
simplices. This procedure in known as Regge discretization. The set of all
Riemannian spaces is then thought of as a limit of the sequence of finite
dimensional ones (each that space corresponds to the given fixed triangulation
with varying link lengths). Then the continuum measure should be the limit of
measures on these finite dimensional approximating spaces. Our aim now is to
construct these finite dimensional measures in such a way that the result
corresponds to the norm (\ref{NORM}).

First of all we note that (\ref{NORM}) induces the so - called Lund - Regge
norm on space of Regge skeletons with fixed triangulation \cite{A1}. To
calculate it explicitly let us equip each simplex $[\Gamma_0 \Gamma_1 ...
\Gamma_D]$ with the basis $\{ \frac{\Gamma_0 \Gamma_1}{|\Gamma_0 \Gamma_1|}$ ,
$\frac{\Gamma_0 \Gamma_2}{|\Gamma_0 \Gamma_2|}$ , ... , $\frac{\Gamma_0
\Gamma_D}{|\Gamma_0 \Gamma_D|} \}$. In this basis metrics is
\begin{equation}
g_{i j} = \frac{1}{2}(|\Gamma_0 \Gamma_i|^2 + |\Gamma_0 \Gamma_j|^2 - |\Gamma_i
\Gamma_j|^2)\label{G}.
\end{equation}
Here $|\Gamma_i \Gamma_j|$ is the distance between points $\Gamma_i$ and
$\Gamma_j$. We can substitute (\ref{G}) into (\ref{NORM}) and obtain the form
quadratic in variations of link lengths (we denote link lengths as $a_i$).

Before doing so let us rewrite (\ref{NORM}) in the following way:
\begin{eqnarray}
\|\delta g\|^2 & = & \frac{1}{2}\int d^D x \sqrt{|g|} (g^{\mu \nu}g^{\rho \eta}
+ g^{\rho \nu}g^{\mu \eta} + C g^{\mu \rho}g^{\nu \eta})\delta g_{\mu \rho}
\delta g_{\nu \eta}\nonumber\\
& = & \int d^D x \sqrt{|g|} ( - \delta g^{\mu \rho} \delta g_{\mu \rho} +
\frac{C}{2} [\delta {\rm log} |g| ]^2)\label{N1}
\end{eqnarray}
We used here, that $\delta |g| = |g| g^{ik} \delta g_{ik}$ and $g_{ji}\delta
g^{ik} = - g^{ik} \delta g_{ji}$.

Let us represent $g^{ij}$ as
\begin{equation}
g^{kl} = \frac{1}{(D-1)!g}\epsilon^{kab...d}\epsilon^{lvg...s} g_{av} g_{bg}
... g_{ds}\label{g}
\end{equation}

Then, we substitute (\ref{g}) into (\ref{N1}) and obtain:
\begin{equation}
\|\delta g\|^2  = \int d^D x  [-
\frac{\epsilon^{kab...d}\epsilon^{lvg...s}\delta g_{kl}\delta g_{av} g_{bg} ...
g_{ds}}{\sqrt{|g|}(D-2)!}
 + \frac{C+2}{2} (\delta {\rm log} |g|)^2 \sqrt{|g|}]\label{N2}
\end{equation}

Now let us introduce new variable ${\hat g}_{ij} = |g|^{-\frac{1}{D}} g_{ij}$.
The determinant $|{\hat g}|$ is equal to unity by construction. The original
field variable $g_{ij}$ is, therefore, represented as a product of this new
variable and the conformal factor $|g|^{\frac{1}{D}}$. As a result (\ref{N2})
can be rewritten as
\begin{equation}
\|\delta g\|^2  = \int d^D x  [-
\frac{\epsilon^{kab...d}\epsilon^{lvg...s}\delta {\hat g}_{kl}\delta {\hat
g}_{av} {\hat g}_{bg} ... {\hat g}_{ds}}{(D-2)!}
 + \frac{C+\frac{2}{D}}{2} (\delta {\rm log} |g|)^2] \sqrt{|g|}\label{N2a}
\end{equation}
Here we can see, that the conformal mode $|g|$ is orthogonal to the field
${\hat g}_{ij}$.

 On the simplicial manyfold expression (\ref{N2}) has the
form:
\begin{equation}
\|\delta a\|^2 = \sum_{\rm simplices} [-
\frac{\epsilon^{kab...d}\epsilon^{lvg...s}\delta g_{kl}\delta g_{av} g_{bg} ...
g_{ds}}{\sqrt{|g|}(D-2)!}
 + \frac{C+2}{2} (\delta {\rm log} |g|)^2 \sqrt{|g|}]
 \label{NORMD2}
\end{equation}
Here\footnote{It is worth mentioning that in a similar way expression
(\ref{N2a}) can be transferred to the simplicial manyfold. However, there the
main property of (\ref{N2a}) is lost. Namely, on the lattice again we may
introduce new variables (instead of the link lengths). These new variables are
triangle areas and the remaining angle variables. Naively one would expect,
that in analogy with ${\hat g}_{ij}$ and $|g|$ the subspaces spanned on these
two sets of variables are orthogonal. But this is not true, because now the
induced ${\hat g}_{ij}$ on the simplices functionally depend both upon
mentioned angle variables and on the areas of the triangles.} at each simplex
$g_{ij}$ is expressed through link lengths $a_m$.

Collecting all terms quadratic in $\delta a_m$ one might obtain the final
result
\begin{equation}
\| \delta a \|^2 = \sum_{i,j \in C_1} O_{i j}[a] \delta (a_i^2) \delta
(a_j^2),\label{na}
\end{equation}
with the matrix $O$, which (in general) depends on link lengths in rather
complicated way. Further we shall show, however, that in two dimensions this
expression can be greatly simplified if we impose rather restrictive but very
natural constraint on the approximating Regge skeleton.

From (\ref{na}) it follows that the resulting measure is given by\footnote{In
this Letter we denote the set of links of the lattice as $C_1$ and the set of
triangles as $C_2$. The number of sites is denoted as $N_0$, the number of
triangles is denoted as $N_2$, the number of links is denoted as $N_1$. }
\cite{W,A1}:
\begin{equation}
D a =  {\rm Det}^{\frac{1}{2}} O[a] \Pi_{i\in C_1}  d a^2_i,\label{DA}
\end{equation}

Expression (\ref{DA}) was called in \cite{A1} De - Witt like measure. However,
we guess it more appropriate to call it Lund - Regge measure, as it is derived
from the expression for metrics, which was given first by Lund and Regge in
their unpublished preprint. As we already told, unfortunately, in general case,
(\ref{DA}) is not suitable for practical calculations due to its nonlocal and
rather complicated form. The main purpose of this Letter is to show that in two
dimensions the considerable simplification may be achieved, which allows to use
Lund - Regge measure in real computer simulations.

So, let us turn to the two - dimensional case. In order to set up the model,
let us first come back to the continuum formulation. Our point is that before
discretizing the theory we perform partial gauge fixing. The gauge condition is
that the conformal mode $|g|$ has a value, which does not depend upon the point
of the surface:
\begin{equation}
\sqrt{|g(x)|} = \sqrt{|g(x_0)|},
\end{equation}
where $g(x)$ is the metric tensor, while $x_0$ is the fixed point of the
surface.  The Faddeev - Popov procedure (when applying to the partition
function of the model with the action $S[g]$) gives
\begin{eqnarray}
Z & = & \int Dg \exp( - S[g])\int D f \delta({\rm log}\,\sqrt{|g^f|} - {\rm log}\, \sqrt{|g^f(x_0)|})\Delta_{FP}[g]\nonumber\\
& = & \int Dg^{f^{-1}} D f \exp( - S[g^{f^{-1}}])  \delta({\rm log}\,\sqrt{|g|} - {\rm log}\,\sqrt{|g(x_0)|}) \Delta_{FP}[g^{f^{-1}}]\nonumber\\
& = & \int Dg \exp( - S[g]) \delta({\rm log}\,\sqrt{|g|} - {\rm log}\,
\sqrt{|g(x_0)|}) \Delta_{FP}[g]
\end{eqnarray}
Here\footnote{It is important to note, that the definition of delta function on
the Riemannian manyfold requires the background metric $g_0$ to be concretized.
Namely, this function can be represented as $\delta(h(x)) = {\rm
lim}_{t\rightarrow \infty} \exp(-t \int h^2(x) \sqrt{g_0(x)} d^2 x)$. In our
formulas we imply that a certain background metric is chosen. And this metric
has nothing to do with our main dynamical variable $g$. (Of course, in the same
way the measure over reparametrizations $D f$ is defined with respect to this
background metrics.)} the reparametrization is denoted as $f$, and its action
on metrics is denoted as $g^f$. We take into account that the action $S[g]$,
the measure $Dg$, and the Faddeev - Popov determinant $\Delta_{FP}[g]$ are
reparametrization invariant.

The Faddeev - Popov determinant is expressed as (the reparametrization is
written here as $ x \rightarrow x - y(x)$) :
\begin{eqnarray}
&& \Delta_{FP}^{-1}[g]|_{\sqrt{|g(x)|} = \sqrt{|g(x_0)|}}  =   \int D y
\delta({\rm log} \, \frac{|g(x - y(x))|^{1/2}(1 +
\frac{1}{2}\partial_{\mu}y^{\mu}(x))}{|g(x_0 - y(x_0))|^{1/2}(1 +
\frac{1}{2}\partial_{\mu}y^{\mu}(x_0))})\nonumber\\
& = & \int D y \delta({\rm log} \, \frac{1 +
\frac{1}{2}\partial_{\mu}y^{\mu}(x)}{1 +
\frac{1}{2}\partial_{\mu}y^{\mu}(x_0)})
 = {\rm  const}
\end{eqnarray}

  Thus,
$\Delta_{FP}(g)$ does not depend upon metrics.

So, we start discretization of quantum gravity from the model with the
partition function
\begin{equation}
Z =  \int Dg \exp( - S[g]) \delta({\rm log}\,\sqrt{|g(x)|} - {\rm log}\,
\sqrt{|g(x_0)|}) \label{CGF}
\end{equation}

In Regge discretization of (\ref{CGF}) we use Lund - Regge measure (\ref{DA})
instead of $Dg$. Instead of the delta function $\delta({\rm log}\,\sqrt{|g(x)|}
- {\rm log}\, \sqrt{|g(x_0)|}) $ we use its discretized version:
\begin{equation}
\Theta[a] = \Pi_{J\in C_2, J \ne J_0} \delta({\rm log} \, A^J - {\rm log}\,
A^{J_0} ),
\end{equation}
where the product is over the triangles (all but the given triangle $J_0$) of
the simplicial manyfold and $A^J$ is the area of the triangle. Equivalently, we
can represent the last expression in the following form:
\begin{equation}
\Theta[a] = \int \Pi_{J\in C_2} \delta({\rm log} \, A^J - {\rm log}\, (V/N_2) )
\frac{d V}{V},
\end{equation}
where $V$ is the overall invariant volume.

So, the partition function of the discretized model is
\begin{equation}
 Z  =   \int Da \exp(-S[a]) \label{CGFD}
\end{equation}
where $S[a]$ is the discretized action and $Da$ is the measure:
\begin{equation}
 Da = \{{\rm Det}^{\frac{1}{2}} O[a]\,  \frac{d
V}{V}\}\, \{\Pi_{i\in C_1} d a^2_i\}  \,\{\Pi_{J\in C_2} \delta({\rm log} \,
A^J - {\rm log}\, (V/N_2) )\} \label{CGFDM}
\end{equation}

In (\ref{CGFDM}) the additional constraint on the link variables has appeared.
We keep areas of all triangles equal to each other. Initially we had $N_1$
degrees of freedom. There are $N_2$ additional conditions. At each site there
are also $2$ local redundant degrees of freedom. It is most simple to
understand their nature if we consider flat surface. Then local movements of
each lattice site within this surface live the approximated manyfold unchanged.
Thus total number of local redundant degrees of freedom is $2 N_0 = N_2 +
4(1-h)$, where $h$ is genus of the surface. For the case of torus the number of
additional conditions is equal to the number of local redundant degrees of
freedom. In case of sphere the number of additional conditions is less than the
number of redundant degrees of freedom. For $h > 1$, however, our constraint
may become too restrictive. Therefore, for the surface of genus $h
> 1$ the number of additional conditions must be reduced.
Further we restrict ourselves with the cases $h = 0,1$.

The main simplification is achieved in the expression for the determinant ${\rm
Det} \, O[a]$  in the special case $C = -2$ when the areas of all triangles are
equal to each other. In this case (\ref{NORMD2}) has the form:
\begin{equation}
\|\delta a\|^2 = - \frac{N_2}{2V} \sum_{J\in C_2}
\epsilon^{ka}\epsilon^{lv}\delta g^J_{kl}\delta g^J_{av}, \label{NORMD2L}
\end{equation}
This expression can be easily rewritten in terms of the link lengths $a_i$:
\begin{equation}
\|\delta a\|^2 = \frac{1}{V}\sum_{ij} U^{ij} \delta (a^2_{i}) \delta (a^2_{j})
\end{equation}
where the matrix $U$ depends upon the triangulation and does not depend on the
link lengths. So, in this case ${\rm Det} \, O[a] \sim  \frac{1}{V^{N_1}}$.

That's why we came to the following

{\bf conclusion:} In two dimensions (for $h = 0, 1; C = -2$) the measure over
discretized geometries \footnote{Here and below we consider measure up to the
factor, which depends  upon the triangulation only.} is
\begin{equation}
D a = \{V^{- 1 + N_1/2} d V\} \, \{ \Pi_{i\in C_1} d [\frac{a^2_i}{V/N_2}]\}\,
\{\Pi_{J\in C_2} \delta({\rm log} \, A^J - {\rm log}\, (V/N_2) )\},\label{DAC2}
\end{equation}

The formula (\ref{DAC2}) can be considered as the main result of this Letter.

Now let us turn to the description of the particular example of the $2D$ model,
which can be investigated in order to examine the measure (\ref{DAC2}). Namely,
we are going to consider the $R^2$ discretized model. Its partition function
has the form:
\begin{equation}
Z = \int D a \exp( - \beta \, \sum_{i\in C_0} \frac{\theta_i^2}{B_i[a]} )
\label{DGF}
\end{equation}
The deficit angle at the point $i$ is denoted as $\theta_i$. Here we use the
traditional definition of discretized squared curvature $\frac{\theta_i}{
B_i[a]}$ (where $B_i[a] = \sum_{i \in J} \frac{1}{3} A^J$ is the sum over
triangles incident at the point $i$).

Now let us consider the partition function for the fixed invariant volume (we
rescaled $a$ as $ a \rightarrow \sqrt{2 V/N_2} a$):
\begin{equation}
Z(V) =  V^{N_1/2-1} \hat{Z}(\hat{V}),
\end{equation}
where
\begin{equation}
\hat{Z}(\hat{V}) =   \int  \Pi_{i\in C_1} d a^2_i \exp( - \frac{1}{\hat{V}}\,
\frac{N_2}{2} \sum_{i\in C_0} \frac{\theta_i^2}{B_i[a]} )\Pi_{J\in
C_2}\delta(2A^J[a]-1), \label{RDGF}
\end{equation}
and $\hat{V} = \frac{V}{\beta}$.

In order to calculate $Z(V)$  and extract the string succeptibility we have to
investigate the correlator
\begin{equation}
F(V,\beta)  =   \frac{\partial \, {\rm log}\, { Z}}{\partial \, {\rm log} { V}}
- (\frac{N_1}{2}-1) = \frac{N_2}{2 {\hat Z} {\hat V}}  < \sum_{i\in C_0}
\frac{\theta_i^2}{B_i[a]}
> \label{CRDGF}
\end{equation}
where the averaging is over the model defined by the partition function
(\ref{RDGF}).

At ${\hat V} >> 1$ the structure of ${ Z}({ V})$ is expected to match the
following expression
\begin{equation}
{ Z}({ V}) \sim { V}^{ - 3 + \gamma} \exp( m  V)), \label{GM}
\end{equation}
where $\gamma$ is the so - called string succeptibility while $m$ is the
renormalized cosmological constant \cite{KPZ}.

On the other hand we expect that in the limit $N_2 >> \hat{V} >> 1$ the system
in leading order behaves like a collection of $N_d$ oscilators, where $N_d$ is
the effective number of degrees of freedom. Thus the leading part of
(\ref{CRDGF}) is
\begin{equation}
F({\hat V}) = \frac{N_d}{2} + o(N_2)
\end{equation}

Naively one would suppose, that $N_d = N_1-N_2$. However, close to the
continuum limit the number of degrees of freedom may be reduced due to the
restored symmetry (which is lost in the discretized model). Nevertheless, we
expect, that this change in the number of degrees of freedoms remains finite at
$N_2 \rightarrow \infty$ because we have already eliminated most of redundant
degrees of freedom via the constraint on the triangle areas. Under this
assumption we can represent $\gamma$ as (we used here, that $N_1 =
\frac{3}{2}N_2$):
\begin{equation}
\gamma =  N_2 + \gamma_{fin} + O(\frac{1}{N_2}), \label{GAMMA}
\end{equation}
where $ \gamma_{fin}$ does not depend upon $N_2$. From (\ref{GAMMA}) it
follows, that, strictly speaking, the string succeptibility is infinite.
Moreover, this statement would remain valid even if the number of degrees of
freedom is reduced by a number, that depends on $N_2$. That's why we came to
the conclusion that value of string succeptibility in our model contradicts
with the one given by the KPZ formula\cite{KPZ}:
\begin{equation}
\gamma = 2 - \frac{5}{2} (1 - h)
\end{equation}

Thus from our analysis it follows that the only possibility to compare KPZ
result with the string succeptibility calculated via the discretized model
given by (\ref{DGF}) is to consider its finite part (which under mentioned
above assumptions can be given, say, by $\gamma_{fin}$ of (\ref{GAMMA})).

It is important to notice that usual Metropolis algorithm applied to the system
(\ref{RDGF}) must be redefined in order to implement the constraints on the
triangle areas. While updating link lengths the condition $A_i = 0.5$ is
resolved as follows. When the link to update is chosen at random, we consider
the two triangles having it as a common side. In each of these two triangles we
choose the vertex, that does not belong to the given link. Next we consider the
set of all triangles having one of these two points as one of their vertices.
The resulting figure consists of two stars with the given  two points as their
centers.

If there are $n_1$ internal links in this figure (the links that do not belong
to the boundary of the figure), then there are either $n_1 - 1$ or $n_1 - 2$
triangles that belong to the figure. So, keeping lengths of boundary links and
the length of the first updated link, in both cases we might be able to resolve
the constraints $A^J = 0.5$ within the figure. However, triangle inequalities
may forbid these constraints to be resolved for certain choices of the first
chosen link length. If so, we change it and repeat resolving constraints (at
least for its initial value they may be resolved by the initial conditions).

In practise we use the constraint
\begin{equation}
|A^J - 0.5| < \delta \label{CON}
\end{equation}
on each triangle. Here $\delta$ is chosen to be sufficiently small. (In
principle it could be made as small as nesesary. However, this may cause
considerable increasing of the computer time. So we must adjust $\delta$ and,
as a result, final accuracy of calculations in order to keep CPU time
acceptable.)

In order to resolve the constraints, we  make several steps over all links of
the constructed figure that are allowed to be changed. Adjusting each link, we
minimize the quantity $(A_1^2 - \frac{1}{4})^2 + (A_2^2 - \frac{1}{4})^2$,
where $A_1$ and $A_2$ are the areas of the triangles having the given link as
common. This minimization is achieved via direct solving of cubic equation. So
this procedure is repeated iteratively until the requirement (\ref{CON}) is
achieved at each triangle.

After this procedure is completed, the suggestion is formed, which includes the
update of the initially chosen link and the calculated update of all other
internal links of the constructed figure. Then, this suggestion is accepted or
rejected with the usual probability ( $ p = 1$ if $\Delta S < 0$, and $ p =
e^{-\Delta S}$ otherwise, where $\Delta S$ is the correspondent change in the
action).

To conclude, we have proposed in this Letter the modification of the $2D$ Regge
calculus, in which the areas of all triangles are equal to each other. As a
result the Lund - Regge measure over link lengths (at $C = -2$) is simplified
considerably and becomes local. The discretized quantum gravity models with
this measure can be investigated numerically in a relatively simple way
contrary to the models with Lund - Regge measure defined on the conventional
Regge lattice. As a result of brief consideration of the squared curvature
model we argue, that the string succeptibility contains an infinite part. And
this is reasonable to compare it with the KPZ result only modulo this infinite
part. We also have described changes in the Metropolis algorithm necessary for
implementing the constraints on the triangle areas.

 It is important to note,
that in order to obtain the mentioned simplification we were forced to start
the discretization of gravity from the partially gauge - fixed version of the
theory. Namely, we used the gauge condition, which makes the conformal mode
$|g|$ of the metric field constant along the surface. The correspondent Faddeev
- Popov determinant is shown to be independent of metric field. So, we lost a
part of reparametrization invariance when coming to our our version of Regge
discretization. This was the price for the simplification of measure. We do not
think this loss to be a pathology, keeping in mind that, say,   Lorentz
invariance is lost in any model defined on rectangular lattice. Nevertheless,
it would be important to check if the mentioned part of reparametrization
invariance is restored or not when a continuum limit of the discretized model
is approached.

The author is grateful to V.Rubakov, E.Akhmedov, A.Smilga, and J. Ambjorn for
useful discussions. He also kindly acknowledges the hospitality of Niels Bohr
Institute at Copenhagen, where this work was initiated. This work was partly
supported by RFBR grants 03-02-16941, 05-02-16306, and 04-02-16079, by Federal
Program of the Russian Ministry of Industry, Science and Technology No
40.052.1.1.1112.

\end{document}